\documentclass[12pt,preprint]{aastex}
\newcommand       \Angstrom     {\,{\rm \AA}} 
          
\newcommand       \cm           {\,{\rm cm}}

\newcommand	  \g		{\,{\rm g}}

\newcommand       \mum          {\,{\rm \mu m}}
\newcommand	  \ppm		{\,{\rm ppm}}

\newcommand	  \NH		{N_{\rm H}}
\newcommand       \simali       {\sim\,}

\newcommand	  \tauext	{\tau_{\rm ext}}
\newcommand	  \Vtot	        {V_{\rm dust}^{\rm tot}}

\newcommand	  \magni	{\,{\rm mag}}

\newcommand	  \xism         {{\small \left[{\rm X/H}\right]_{\rm tot}}}
\newcommand	  \csun         {{\small \left[{\rm C/H}\right]_{\odot}}}

\newcommand	  \fesun        {{\small \left[{\rm Fe/H}\right]_{\odot}}}

\newcommand	  \xdust        {{\small \left[{\rm X/H}\right]_{\rm dust}}}

\newcommand	  \xgas         {{\small \left[{\rm X/H}\right]_{\rm gas}}}

\newcommand \fesk    {{\small \left[{\rm Fe/H}\right]^{\rm Sk\,155}_{\rm tot}}}
\newcommand \znsk    {{\small \left[{\rm Zn/H}\right]^{\rm Sk\,155}_{\rm tot}}}
\newcommand \znsun   {{\small \left[{\rm Zn/H}\right]_{\odot}}}
\newcommand \csk     {{\small \left[{\rm C/H}\right]^{\rm Sk\,155}_{\rm tot}}}
\newcommand \ssk     {{\small \left[{\rm S/H}\right]^{\rm Sk\,155}_{\rm tot}}}

\newcommand	  \mux         {\mu_{\rm X}}

\newcommand \FPmin {\small \left[F/\left(1-P\right)\right]^{\rm obs}_{\rm min}}
\newcommand \FPmod {\small \left[F/\left(1-P\right)\right]_{\rm mod}}
\newcommand	  \fmin  {f_{\rm obs}^{\rm min}}
\newcommand	  \fmod  {f_{\rm mod}}
\shorttitle{Unusual SMC Depletions}

\begin{document}

\title{
 \vspace*{-2.0em}
  {\normalsize\rm {\it The Astrophysical Journal Letters}, in press}\\
 \vspace*{1.0em}
On the Unusual Depletions toward Sk 155
or What Are the Small Magellanic Cloud Dust Grains Made of?
	 }
\author{Aigen Li\altaffilmark{1}, K.A. Misselt\altaffilmark{2}, 
and Y.J. Wang\altaffilmark{3}}
\altaffiltext{1}{Department of Physics and Astronomy,
                 University of Missouri, Columbia, MO 65211;
                 {\sf LiA@missouri.edu}}
\altaffiltext{2}{Steward Observatory,
                 University of Arizona, Tucson, AZ 85721;
                 {\sf kmisselt@as.arizona.edu}}
\altaffiltext{3}{Department of Physics, Hunan Normal University,
                 Changsha 410071, P.R.~China;
                 {\sf wyj@hunnu.edu.cn}}

\begin{abstract}
The dust in the Small Magellanic Cloud (SMC), an ideal analog 
of primordial galaxies at high redshifts, differs markedly from 
that in the Milky Way by exhibiting a steeply rising 
far-ultraviolet extinction curve,
an absence of the 2175$\Angstrom$ extinction feature,
and a local minimum at $\simali$12$\mum$ in its infrared 
emission spectrum, suggesting the lack of ultrasmall carbonaceous 
grains (i.e. polycyclic aromatic hydrocarbon molecules) 
which are ubiquitously seen in the Milky Way. 
While current models for the SMC dust all rely heavily on
silicates, recent observations of the SMC sightline toward Sk\,155 
indicated that Si and Mg are essentially undepleted 
and the depletions of Fe range from mild to severe, 
suggesting that metallic grains and/or iron oxides, 
instead of silicates, may dominate the SMC dust. 
However, in this {\it Letter} we apply the Kramers-Kronig 
relation to demonstrate that neither metallic grains 
nor iron oxides are capable of accounting for 
the observed extinction; 
silicates remain as
an important contributor to the extinction, 
consistent with current models for the SMC dust. 

\end{abstract}

\section{Introduction\label{sec:intro}}
As a metal-poor (with a metallicity only $\sim$1/10 of 
that in the Milky Way; see Kurt \& Dufour 1998)
and gas-rich (with a dust-to-gas ratio over 10 times 
lower than in the Milky Way; see Bouchet et al.\ 1985)
irregular dwarf galaxy, the Small Magellanic Cloud (SMC)
is often considered as a local analog of primordial galaxies 
at high redshifts which must have formed at very low metallicity.
Therefore, the dust in the SMC which differs substantially from
that in the Milky Way (see Li \& Draine 2002) allows us to probe 
the primordial conditions in more distant galaxies.

The SMC extinction curve 
displays a nearly linear rise with
inverse wavelength
and no detectable 2175$\Angstrom$ extinction bump
(
Lequeux et al.\ 1982; 
Cartledge et al.\ 2005),
presumably due to the destruction of the carriers
of the 2175\AA\ hump by the intense ultraviolet (UV)
radiation and shocks associated with star formation
[an exception to this is the line of sight toward  
  Sk 143 (AvZ 456) for which the extinction curve has a strong 
  2175\AA\ hump (Lequeux et al.\ 1982; 
  Cartledge et al.\ 2005). 
  This sightline passes through the SMC wing,
  a region with much weaker star formation (Gordon \& Clayton 1998)]. 
Although the precise nature of the carriers of the 2175\AA\ hump  
remains unclear, it is generally believed to be due to 
the $\pi$$\rightarrow$$\pi^{\ast}$ transition of small aromatic 
carbonaceous (i.e. graphitic) grains, 
probably a cosmic mixture of polycyclic aromatic hydrocarbon 
(PAH) molecules (Li \& Draine 2001).
%
The overall infrared (IR) emission spectrum of the SMC
peaks at $\lambda$$\simali$100$\mum$ 
with a local minimum at $\lambda$$\simali$12$\mum$
which is commonly believed to be emitted by PAHs 
(see Stanimirovic et al.\ 2000, Li \& Draine 2002 and references therein).

The absence of the 2175$\Angstrom$ extinction bump
and the very weak 12$\mum$ emission in the SMC imply 
the lack of PAHs in the SMC.\footnote{%
  The PAH emission features have been detected locally 
  -- in the SMC B1 No.\,1 quiescent molecular cloud
  (Reach et al.\ 2000), and in the N\,66 star-forming region
  (Contursi et al.\ 2000). 
  }
The paucity of PAHs appears to be a general feature 
of metal-poor galaxies (e.g. see Madden 2000, 
Thuan et al.\ 1999, Houck et al.\ 2004, 
Engelbracht et al.\ 2005, Wu et al.\ 2006).
A natural question one may ask is: 
are their large-size counterparts 
-- carbon dust (either graphite or amorphous carbon) with 
radii larger than $\simali$100$\Angstrom$ -- also deficient
in the SMC and low-metallicity galaxies, or more generally,
what are the dust grains in the SMC 
(or metal-poor galaxies) made of?

Our knowledge about the SMC dust is mainly derived from
the extinction curve. As early as 1983 when 
the $\simali$3--9$\mum^{-1}$ UV extinction data were just 
available for the SMC 
thanks to the IUE satellite,
Bromage \& Nandy (1983) 
had already recognized ``the conspicuous {\it absence} of 
normal {\it graphite} grains in the SMC''. 
Subsequent models for the SMC dust 
all rely heavily on {\it silicates}, 
with a silicate to carbon mass ratio 
of $\simali$5.5 (Rodrigues et al.\ 1997),
$\simali$2.5 (Zubko 1999),
$\simali$12 (Weingartner \& Draine 2001), 
$\simali$1.6 (Clayton et al.\ 2003),
and $\simali$12 (Cartledge et al.\ 2005).
Pei (1992) even attributed the SMC extinction 
exclusively to silicates alone.
However, based on an analysis of the HST/STIS interstellar 
absorption spectra of the SMC components,
Welty et al.\ (2001) recently reported that 
the Si and Mg elements in the interstellar cloud toward 
Sk\,155 in the SMC appear to be essentially {\it undepleted}, 
imposing a serious {\it challenge} on all dust models 
for the SMC since they all require a substantial amount of
silicates to account for the observed extinction.

It is the purpose of this {\it Letter} to examine
the {\it unusual} depletions observed in the SMC
sightline toward Sk\,155. To this end, we investigate
whether the depleted elements are sufficient to account for
the observed extinction. Our approach, 
as described in \S\ref{sec:kk}, is based on
the Kramers-Kronig (KK) dispersion relation (Purcell 1969)
and is independent of any specific dust models.

\section{Constraints from the Kramers-Kronig Relation\label{sec:kk}}
In 1926--1927, H.A. Kramers and R. Kronig independently demonstrated
that the real (dispersive) and imaginary (absorptive) parts 
of the index of refraction are not independent, 
but are related to each other through a relation which is now known 
as the Kramers-Kronig (KK) dispersion relation. 
Applying this relation to the ISM, Purcell (1969) related 
the interstellar extinction integrated over wavelength 
to the total volume of grains in the ISM: 
$\int_{0}^{\infty} \tauext(\lambda)/N_{\rm H}\,d\lambda
= 3 \pi^2 F\,\Vtot/{\rm H}$,
where 
$\tauext(\lambda)$ is the extinction optical depth
at wavelength $\lambda$, 
$\NH$ is the H column density,
$\Vtot/{\rm H}$ is the total volume occupied by dust 
per H nucleon, 
and the dimensionless factor $F$ 
is the orientationally-averaged polarizability 
relative to the polarizability of an equal-volume 
conducting sphere, 
which depends only upon 
the grain shape and the static (zero-frequency) 
dielectric constant $\epsilon_0$ of the grain 
material (Purcell 1969).\footnote{%
  The static dielectric constants of the dust materials 
  adopted in this {\it Letter} are:
  $\epsilon_0\approx 10$ for amorphous olivine (Draine \& Lee 1984), 
  $\epsilon_0\approx 9$ for MgO (Roessler \& Huffman 1998), 
  $\epsilon_0\approx 28$ for FeO, 
  $\epsilon_0\approx 16$ for Fe$_2$O$_3$ (Steyer 1974), 
  $\epsilon_0\approx 23$ for Fe$_3$O$_4$ (Landolt \& B\"ornstein Tables), 
  and $\epsilon_0\approx 6$ for amorphous carbon 
  (J. Robertson 2005, private communication).
  }

Very recently, we have applied this relation to examine
whether porous dust could solve the so-called interstellar 
subsolar abundance ``crisis'' (Li 2005).
Similarly, this approach can be applied to the unusual
depletions seen along the Sk\,155 sightline: if Si and Mg
are indeed undepleted, would there be enough raw material 
to form the dust to account for the extinction observed
for the SMC? 

To make an economical use of the heavy elements,
we consider fluffy dust with a ``fluffiness'' or ``porosity'' 
of $P$ (i.e. the volume fraction of vacuum contained in a grain).
Assuming that all heavy elements depleted from the gas phase
have been locked up in dust which consists of $N$ grain species,
the {\it maximum} volume of dust per H nucleon 
can be estimated from 
$\Vtot/{\rm H}$\,=\,$\sum_{j=1}^{N}\sum_{\rm X} 
f_{{\rm X},j} \left(\xism-\xgas\right) 
\mu_{\rm X}/\left(1-P\right) \rho_j$,
where 
$\xism$ and $\xgas$ are respectively the total and gas abundance 
of element X relative to H (the amount of X contained in dust 
is therefore $\xdust$\,=\,$\xism-\xgas$);
$\mux$ is the atomic weight of X;
$\rho_j$ is the mass density of dust species $j$
(we take $\rho$\,=\,3.5, 3.58, 5.7, 5.25, 5.18, 1.8, 
and 7.85$\g\cm^{-3}$ for amorphous olivine MgFeSiO$_4$, 
MgO, FeO, Fe$_2$O$_3$, Fe$_3$O$_4$, amorphous carbon, 
and pure iron grains, respectively); 
and $f_{{\rm X},j}$ is the fraction of element X
locked up in dust species $j$. 
The first summation is over all possible 
dust species and the second one is over all condensible elements.
Since $\tauext(\lambda)$ is a positive number for all wavelengths,
we can use the extinction observed over a {\it finite} wavelength 
range (say, from 912$\Angstrom$ to 1000$\mu$m) 
to place a lower bound on $F/\left(1-P\right)$:
$f_{\rm obs}^{\rm min}$\,$\equiv$ 
${\small \left[F/\left(1-P\right)\right]^{\rm obs}_{\rm min}}$
=\,$\int_{912\Angstrom}^{1000\mum} 
\tauext(\lambda)/\NH\,d\lambda/\left\{ 
3 \pi^2 \sum_{j=1}^{N}\sum_{\rm X} 
f_{{\rm X},j} \left(\xism-\xgas\right) 
\mu_{\rm X}/\rho_j\right\}$.

The SMC selective extinction toward the Sk\,155 sightline
is $E(B-V)\approx 0.09\magni$ (Fitzpatrick 1985).
From fitting the STIS Ly$\alpha$ data
and the FUSE Ly$\beta$ data, 
D.E. Welty (2005, private communication) determined
the HI column density to be 
$N_{\rm HI}\approx 2.51\times 10^{21}\cm^{-2}$
for this region;
he also estimated the H$_2$ column density from FUSE
data to be $N_{\rm H_2}\approx 1.38\times 10^{19}\cm^{-2}$.
With a total-to-selective extinction ratio of 
$R_V\approx 2.1$, typical for the SMC wing
  where Sk\,155 is located (Clayton et al.\ 2003),
the visual extinction per H column for Sk\,155 is 
$A_V/\NH = R_V\,E(B-V)/\left(N_{\rm HI}+2\,N_{\rm H_2}\right)
\approx 7.44\times 10^{-23}\magni\cm^2$.
For $912\Angstrom$\,$<$\,$\lambda$\,$<$\,3$\mum$, 
we take the extinction curve typical for the SMC bar
(Gordon \& Clayton 1998) for Sk\,155 
since Preliminary analysis of the extinction toward Sk\,155
appears that it is more similar to the ``typical'' SMC 
bar curve than to the curve for Sk\,143
(D.E. Welty 2005, private communication);
for $3\mum$\,$<$\,$\lambda$\,$<$\,1000$\mum$
we adopt the theoretical $\tauext(\lambda)/\NH$ values
calculated from the silicate-graphite-PAH model for
the SMC bar which has been shown to successfully reproduce 
the observed extinction from the far-UV to mid-IR and 
the observed IR emission
(Weingartner \& Draine 2001, Li \& Draine 2002).
The integration of the extinction over 
the 912$\Angstrom$--1000$\mum$ wavelength range 
is approximately 
$\int_{912\Angstrom}^{1000\mum} \tauext(\lambda)/\NH\,d\lambda
\approx 2.08\times 10^{-26}\cm^{3}/{\rm H}$.

If we know the total abundances $\xism$ and gas-phase 
abundances $\xgas$ for the key dust-forming heavy elements,
we can derive $\fmin$ from above through a general assumption 
of the constituent dust materials (e.g. silicates, amorphous 
carbon, magnesium oxides, or iron oxides)
of the porous composite grain with a given porosity $P$,
without any detailed knowledge of the dust sizes and shapes. 
On the other hand, for such a grain with 
a given porosity $P$ and a given shape (e.g. prolate or oblate), 
we can calculate the theoretical values of 
$\fmod$\,$\equiv$\,$\FPmod$ 
(see Purcell 1969, Li 2005).
Apparently, for any valid grain models, the model-predicted
$\fmod$ should exceed $\fmin$.
Below we examine the 2 possible grain models implied from
the anomalous depletions suggested by Welty et al.\ (2001). 

\begin{figure}
\plotone{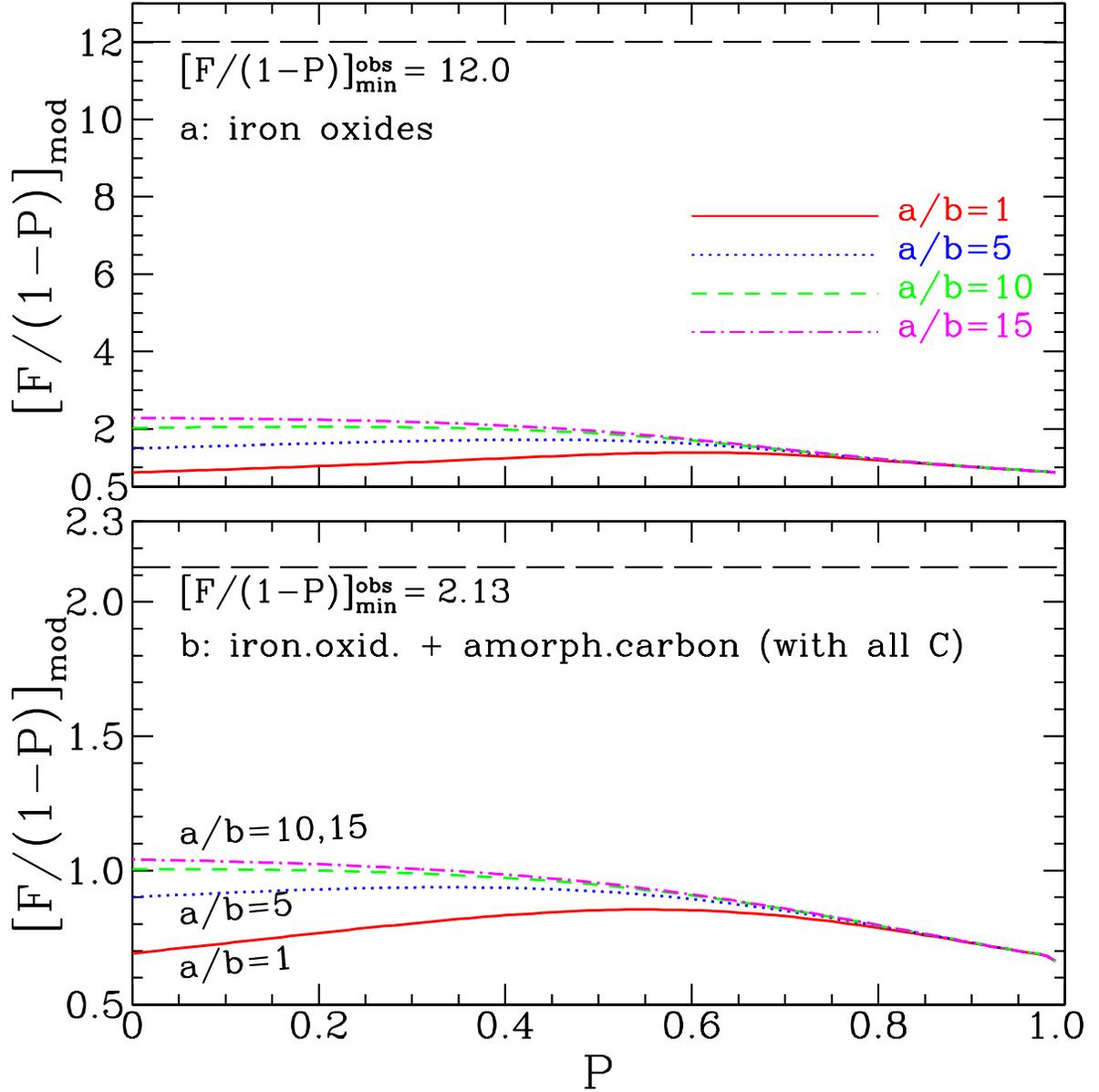}
\caption{
        \label{fig:iron}
        Model-predicted $\fmod$\,$\equiv$\,$\FPmod$ 
        as a function of porosity $P$
        for spherical (solid lines) or prolate porous composite
        grains consisting of 
        (a) a mixture of iron oxides 
        (FeO, Fe$_2$O$_3$, and Fe$_3$O$_4$),
        or (b) a mixture of FeO, Fe$_2$O$_3$, Fe$_3$O$_4$
        and amorphous carbon             
        with an elongation of $a/b=5$ (dotted lines),
        $a/b=10$ (dashed lines), and $a/b=15$ (dot-dashed lines).
        The Fe and C abundances are taken to be solar scaled by Zn 
        (Welty et al.\ 2001).
        The horizontal long-dashed lines plot 
        the lower boundary $\fmin$\,$\equiv$\,$\FPmin$ implied from
        the SMC extinction.
        }
\end{figure}

(1) {\it Metallic grains}? --- Welty et al.\ (2001) suggested
that metallic grains or iron oxides may dominate the dust
populations in the SMC sightline toward Sk\,155 since they
thought that Si and Mg are essentially undepleted while Fe
is severely depleted.
If we take the abundances of heavy elements 
to be those of solar
(
Asplund, Grevesse, \& Sauval 2005)
scaled by the abundance of Zn 
($\znsk\approx 0.004$\,ppm; Welty et al.\ 2001)
which is considered typically undepleted, 
the total iron abundance is
$\fesk = \znsk \fesun/\znsun \approx 2.5\ppm$. 
For compact metallic iron grains
($P=0$), this implies $\fmin\approx 24$.
In order to have $\fmod > 24$, 
these grains need to be highly elongated: 
with an elongation of at least $\simali$25
for prolates and $\simali$84 for oblates 
(see Fig.\,3 of Li 2003b).\footnote{%
  We do not consider {\it porous} metallic needles since it is
  hard for us to believe that they could form and survive 
  in the SMC. 
  }
However, these needle-like iron grains 
would produce an extinction curve very different from 
that observed for the SMC
since their extinction cross sections 
are essentially constant at wavelengths shorter than 
the long-wavelength cutoff and beyond which they decline 
as $\lambda^{-2}$ (see Eqs.[3-5] of Li 2003a).
Therefore, metallic iron grains are unlikely the dominant dust 
component in the SMC.

(2) {\it Iron oxides}? --- Assuming that all 2.5$\ppm$ Fe 
(relative to H) are evenly tied up in FeO, Fe$_2$O$_3$,
and Fe$_3$O$_4$, porous grains consisting of FeO, Fe$_2$O$_3$,
and Fe$_3$O$_4$ would have $\fmin\approx 12$.
However, as shown in Figure \ref{fig:iron}a, 
the $\fmod$ values predicted for the porous 
iron oxide dust are much smaller than $\fmin$ 
even for grains with an elongation
as large as 15.
Therefore, iron oxides are unlikely 
the dominant dust components in the SMC.
Further more, even if we assume that 
(1) there is another major dust species
-- amorphous carbon,\footnote{%
  We do not consider graphite grains because if
  they are the dominant contributor to the SMC extinction,
  they need to be very small; but very small graphite grains
  would produce a strong 2175$\Angstrom$ extinction bump which
  is not commonly seen in the SMC. 
  }
(2) the total C abundance is that of solar
scaled by the abundance of Zn
(i.e. $\csk = \znsk \csun/\znsun \approx 24.5\ppm$), 
and (3) {\it all} C atoms are tied up in amorphous carbon grains,
the model-predicted $\fmod$ values 
are still much smaller than the lower-limit of
$\fmin\approx 2.13$ (see Fig.\,\ref{fig:iron}b). 
Therefore, the extinction observed for the Sk\,155 sightline
is unlikely produced by a mixture of 
iron oxides and amorphous carbon.

\begin{figure}
\plotone{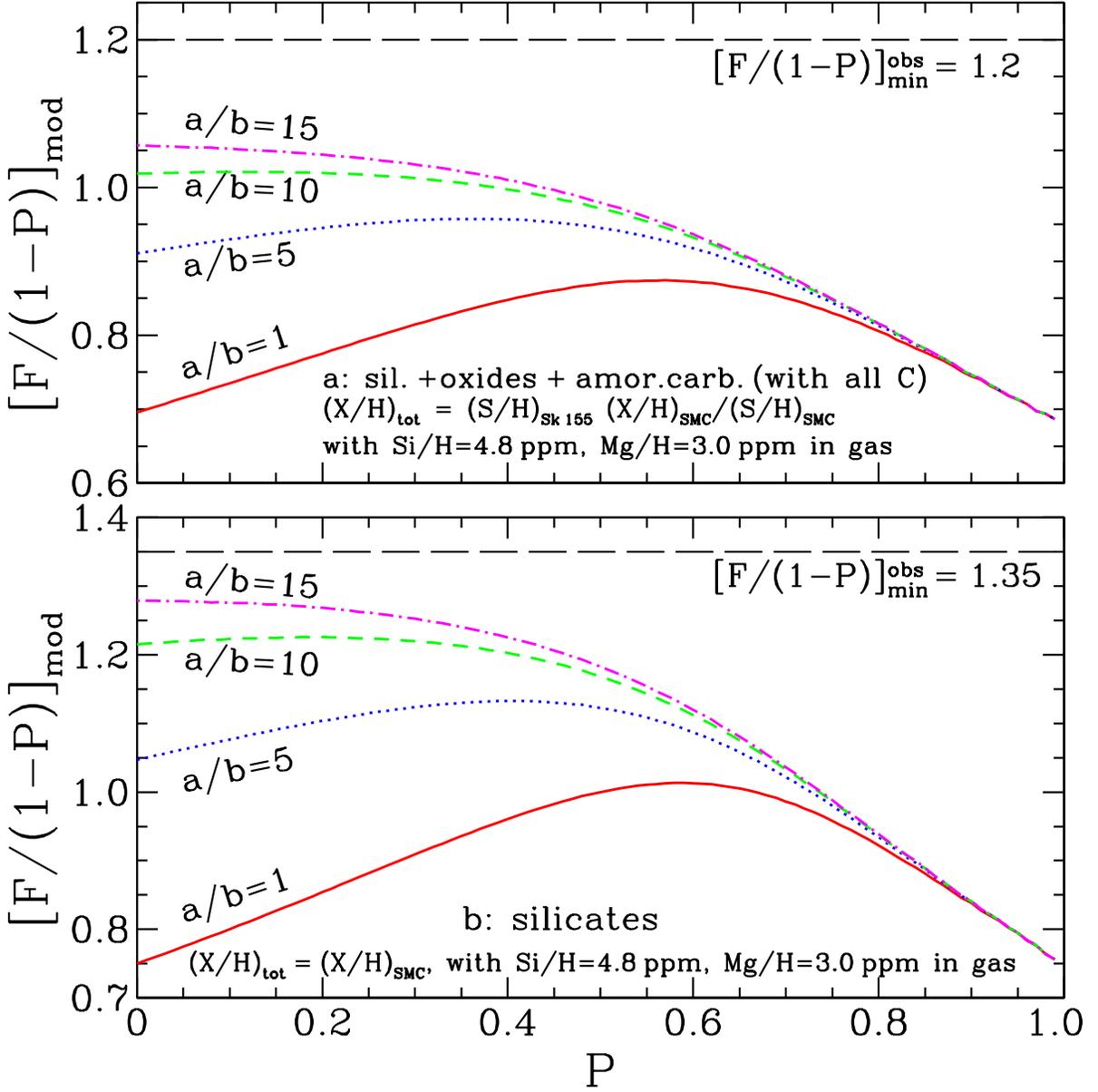}
\caption{
        \label{fig:S}
        Same as Figure \ref{fig:iron}
        but for (a) models consisting of a mixture of 
        silicates, iron oxides, and amorphous carbon
        for which the total abundances are taken to be
        those of the RD92 SMC
        but scaled by the S abundance;
        and (b) models consisting of only silicate dust 
        for which the RD92 SMC abundances
        are adopted as the total abundances. 
        For both models, the Si and Mg gas-phase abundances
        of Welty et al.\ (2001) for Sk\,155 are subtracted.
        }
\end{figure}

\section{Discussion\label{sec:discussion}}

The anomalous depletions in the SMC sightline toward Sk\,155
was derived by comparing the relative abundances [X/Zn] 
in the SMC with the corresponding patterns seen in 
the Galactic ISM, and assuming the relative total 
abundances of the SMC are not very different from 
those of solar (Welty et al.\ 2001).
Vladilo (2002) argued that the SMC abundances
may deviate from the solar ratios.
Sofia et al.\ (2006) argued that Zn is not a suitable
reference for comparison since Zn is modestly depleted 
in the Milky Way ISM;
they further suggested that S may be 
a better comparison species 
since (1) S is essentially undepleted,
and (2) both Si and S are formed through the same
process (i.e. oxygen burning). 
 
With S as the comparison species and adopting the solar
abundances but scaled by the Sk\,155 S abundance
($\ssk \approx 2.4\ppm$, Welty et al.\ 2001), 
we find that the available heavy elements are not sufficient 
to account for the observed extinction. One may argue that
it is more appropriate to use the SMC abundances 
(Russell \& Dopita 1992; hereafter RD92) as a reference rather 
than those of solar (i.e. assuming the abundances for the Sk\,155 
sightline to be those of the SMC scaled by S).
In this case, the total abundances (relative to H) 
of the key dust-forming elements would be
${\rm \{Si, Mg, Fe, C\}} \approx 
\{6.6, 5.9, 4.3, 33\,{\rm ppm}\}$.
Subtracting the gas-phase abundances of
${\rm \{Si, Mg, Fe\}} \approx 
\{4.8, 3.0, 0.35\,{\rm ppm}\}$ of Sk\,155 (Welty et al.\ 2001),
the abundances available for depletion are
${\rm \{Si, Mg, Fe\}} \approx \{1.8, 2.9, 3.9\,{\rm ppm}\}$.
We assume that all the dust-phase Si atoms
are incorporated into amorphous olivine MgFeSiO$_4$.
This will also consume 1.8$\ppm$ Mg and Fe.   
We take the remaining 1.1$\ppm$ Mg to be depleted
in MgO, and the remaining 2.1$\ppm$ Fe evenly tied
up in FeO, Fe$_2$O$_3$, and Fe$_3$O$_4$. 
We also assume that {\it all} the 33$\ppm$ C are bound up 
in amorphous carbon. 
In so doing, we obtain $\fmin \approx 1.2$.
However, as shown in Figure \ref{fig:S}a,
the model-predicted $\fmod$ values never exceed
$\fmin$ even for grains with rather large elongations.
Therefore, the total abundances for Sk\,155 are unlikely
just the RD92 SMC abundances scaled by S.

\begin{figure}
\plotone{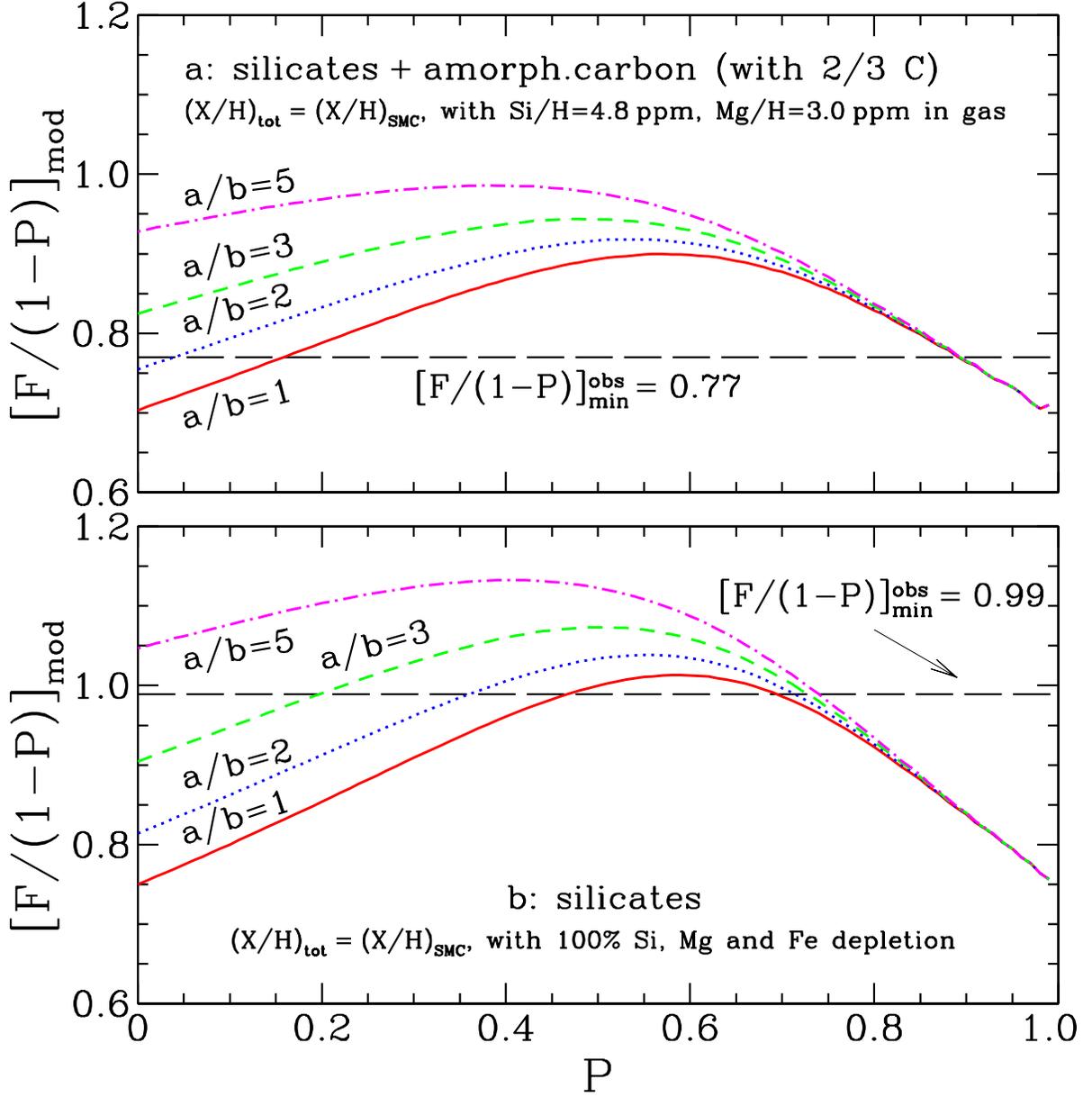}
\caption{
        \label{fig:Si}
        Same as Figure \ref{fig:iron}     
        but for (a) models consisting of
        a mixture of silicates and amorphous carbon
        (with a more modest elongation) 
        for which the total abundances are taken to
        be those of the RD92 SMC
        with the Si and Mg gas-phase abundances of
        Welty et al.\ (2001) subtracted
        and with 2/3 of the total C locked up in amorphous carbon;
        and (b) models consisting of only silicates
        for which the dust-phase Si, Mg and Fe abundances
        are taken to be those of the RD92 SMC abundances
        (i.e. assuming 100\% depletions of Si, Mg and Fe).
        }
\end{figure}

If we simply adopt the RD92 SMC abundances 
as the total abundances for the Sk\,155 sightline
(i.e. without scaling down the SMC abundances), 
the dust-phase abundances would be
${\rm \{Si, Mg, Fe\}} \approx 
\{5.9, 6.6, 6.6\,{\rm ppm}\}$ 
after subtracting the gas-phase abundances
of Welty et al.\ (2001). 
This depletion pattern roughly points to 
a composition of olivine silicates. 
As shown in Figure \ref{fig:S}b,
the $\fmod$ values calculated from these
silicate grains are still smaller than the observational
lower boundary of $\fmin\approx 1.35$ even for grains
with an elongation as large as 15.\footnote{%
  With uncertainties as large as $\simali$20\%
  for $A_V/N_{\rm H}$, silicate grains (together
  with oxides and/or amorphous carbon) with large
  elongations can account for the observed extinction
  (see Figs.\,2a,b), but oxides (even together with
  amorphous carbon) are unable to do it (see Figs.\,1a,b).   
  }
This suggests that either the actual abundances for 
the heavy elements Mg, Si and Fe toward Sk\,155 are
higher than those of RD92, 
or alternatively, there must exist another dust component 
(e.g. amorphous carbon) which makes a considerable 
contribution to the extinction.
For the latter case, with the addition of an amorphous carbon
component which takes 2/3 of the total SMC C abundance
(like the Milky Way diffuse ISM in which about 2/3 of the total 
C are depleted to form carbon dust; see Li 2005),
the model-predicted $\fmod$ values exceed $\fmin\approx 0.77$ 
for grains of any shape and of a wide range of porosities 
(see Fig.\,\ref{fig:Si}a). 
This implies that grain models consisting of a mixture of
silicate and carbon dust are capable of accounting for
the SMC extinction.

The calculations discussed above all put a substantial
fraction of Si and Mg in gas, as found for the Sk\,155 
sightline by Welty et al.\ (2001).
Existing models for the SMC dust all assume a nearly
complete depletion of Si, Mg and Fe into silicates.
As shown in Figure\,\ref{fig:Si}b, the $\fmod$ values 
calculated from silicates consuming all of the Si, Mg, 
and Fe elements (i.e. no gas-phase Si, Mg and Fe)
of the RD92 SMC abundances exceed 
$\fmin\approx 0.99$ for grains with modest elongations 
and a wide range of porosities 
(the Sk\,155 dust is expected to be elongated 
since an appreciable amount of polarization has been detected 
along the Sk\,155 sightline; see Wayte 1990).\footnote{%
  Figure \ref{fig:Si}b shows that with the RD92
  SMC abundances, {\it compact spherical} silicate grains 
  are not able to account for the SMC extinction.
  The reason why Pei (1992) could fit the SMC extinction 
  curve using such grains was because he adopted 
  a higher Si abundance (higher than the RD92
  SMC abundance by $\simali$30\%). 
  In the Weingartner \& Draine (2001) model for the SMC,
  the extinction is also dominantly contributed by 
  compact spherical silicate grains.
  This was also because they took a Si abundance $\simali$37\%
  higher than that of RD92.  
  } 
On the other hand, even if we assume that all the C elements 
are locked up in amorphous carbon, they alone are not able to 
account for the extinction since the model-predicted $\fmod$ 
values are always smaller than $\fmin\approx 1.2$,
suggesting that carbon dust can not be the dominant grain
component in Sk\,155.

\section{Summary\label{sec:summary}}
By comparing the abundances of heavy elements
relative to Zn in the SMC with the corresponding 
patterns seen in the Galactic ISM, and assuming 
the relative total abundances of the SMC are not 
very different from those of solar,
Welty et al.\ (2001) derived the undepletion of Si and Mg 
and mild to severe depletions of Fe for the SMC star Sk\,155.
They further suggested that iron oxides and/or metallic grains,
instead of silicates, may dominate the SMC dust, 
in marked contrast with current dust models for the SMC
which all rely heavily on silicates.
Based on the Kramers-Kronig relation, we study these anomalous 
depletions and their implications for our understanding of 
the SMC dust. It is found that neither iron oxides nor metallic 
grains (even with the addition of an amorphous carbon dust component) 
can account for the observed extinction.
If using S as a reference and scaling either the solar abundances 
or the SMC abundances, the resulting abundances of the condensible
elements are also insufficient to explain the observed extinction. 
However, we are able to account for the SMC extinction 
if adopting the SMC abundances and assuming all Si, Mg and Fe elements 
are locked up in silicates
or a combination of a partial depletion of these elements in silicates 
and a partial depletion of C in amorphous carbon. 
In both cases, silicates are a major contributor to the SMC extinction.

\acknowledgments We thank D.E. Welty and the anonymous referee
for their very helpful comments. This work is in part supported 
by the University of Missouri Summer Research Fellowship, 
the University of Missouri Research Board, and the NASA award P20436.

\end{document}